\documentclass[sigconf]{acmart}

\makeatletter
\def\subsubsection{\@startsection{subsubsection}{3}%
  \z@{.5\linespacing\@plus.7\linespacing}{.1\linespacing}%
  {\normalfont\itshape}}
\makeatother

\usepackage[parfill]{parskip}
\usepackage[algoruled,boxed,lined]{algorithm2e}
\usepackage{amsmath}
\usepackage{amsfonts}
\usepackage{graphicx}
\usepackage{stfloats}
\usepackage{subcaption}
\usepackage{balance}
\usepackage[shortlabels]{enumitem}
\usepackage{url}
\usepackage{listings}
\usepackage{spreadtab}

\setlist[enumerate]{nosep}
\newcommand{\titlefmt}{Input-Aware Auto-Tuning \\ of Compute-Bound HPC Kernels}
\newcommand{\titlenofmt}{Input-Aware Auto-Tuning of Compute-Bound HPC Kernels}

\begin{document}
\copyrightyear{2017} 
\acmYear{2017} 
\setcopyright{acmcopyright}
\acmConference{SC17}{November 12--17, 2017}{Denver, CO, USA}\acmPrice{15.00}\acmDOI{10.1145/3126908.3126939}
\acmISBN{978-1-4503-5114-0/17/11}

\graphicspath{{./figures/}}

\makeatletter
\def\@copyrightspace{\relax}
\makeatother

\title{\titlefmt}
\newcommand{\vx}{\mathbf{x}}
\newcommand{\vy}{\mathbf{y}}
\newcommand{\vz}{\mathbf{z}}
\newcommand{\va}{\mathbf{a}}
\newcommand{\setX}{\mathcal{X}}
\newcommand{\setY}{\mathcal{Y}}
\newcommand{\ts}{\textsuperscript}
\newcommand{\setR}{\mathcal{R}}
\newcommand{\setC}{\mathcal{C}}	
\newcommand\numberthis{\addtocounter{equation}{1}\tag{\theequation}}

\author{Philippe Tillet}
\affiliation{\institution{Harvard University}}
\email{ptillet@g.harvard.edu}

\author{David Cox}
\affiliation{\institution{Harvard University}}
\email{davidcox@fas.harvard.edu}

\begin{abstract}
Efficient implementations of HPC applications for parallel architectures generally rely on external software packages (e.g., BLAS, LAPACK, CUDNN). While these libraries provide highly optimized routines for certain characteristics of inputs (e.g., square matrices), they generally do not retain optimal performance across the wide range of problems encountered in practice. In this paper, we present an input-aware auto-tuning framework for matrix multiplications and convolutions, ISAAC, which uses predictive modeling techniques to drive highly parameterized PTX code templates towards not only hardware-, but also application-specific kernels. Numerical experiments on the NVIDIA Maxwell and Pascal architectures show up to 3x performance gains over both cuBLAS and cuDNN after only a few hours of auto-tuning.

\end{abstract}

\maketitle
\title{\titlenofmt}

\section{Introduction}
The growing adoption of many-core devices across HPC applications has rendered on-node performance perhaps more important than ever. However, while many practitioners have effectively been able to offload the execution of compute- or data-intensive tasks to local accelerators, the wide variety of applications and  architectures available on the market  has made it increasingly challenging to write code whose performance is portable.

In order to develop efficient application code for these diverse architectures, many developers have relied, directly or through external libraries, on automatic source code tuning (auto-tuning). There, the performance-critical portions (kernels) of the application code are parameterized, and those parameters optimized for the architecture -- and inputs -- of interest \cite{fursin2002, yotov2003}. The wide adoption of this technique in fields like Linear Algebra \cite{whaley98, wang2013, tillet2012} and Machine Learning \cite{xiong2001,bergstra2012} has given rise to a plethora of hardware-oblivious software libraries capable of efficiently adapting virtually any underlying memory hierarchies and/or multi-threading schemes.

The resulting performance gains have nonetheless remained highly input\footnote{By \emph{input} we refer to the \emph{characteristics} (matrix dimensions, transposition layout, data-type, etc.) of the input in question rather than the data itself.}-sensitive, often lacking portability across the wide range of problem characteristics encountered in practice; it is for instance common for Basic Linear Algebra Subroutines (BLAS) to be used for computations involving input matrices of certain aspect ratios beyond those for which the implementation was optimized (usually square or highly rectangular).

\begin{figure}[h!]
\centering
\includegraphics[scale=.3]{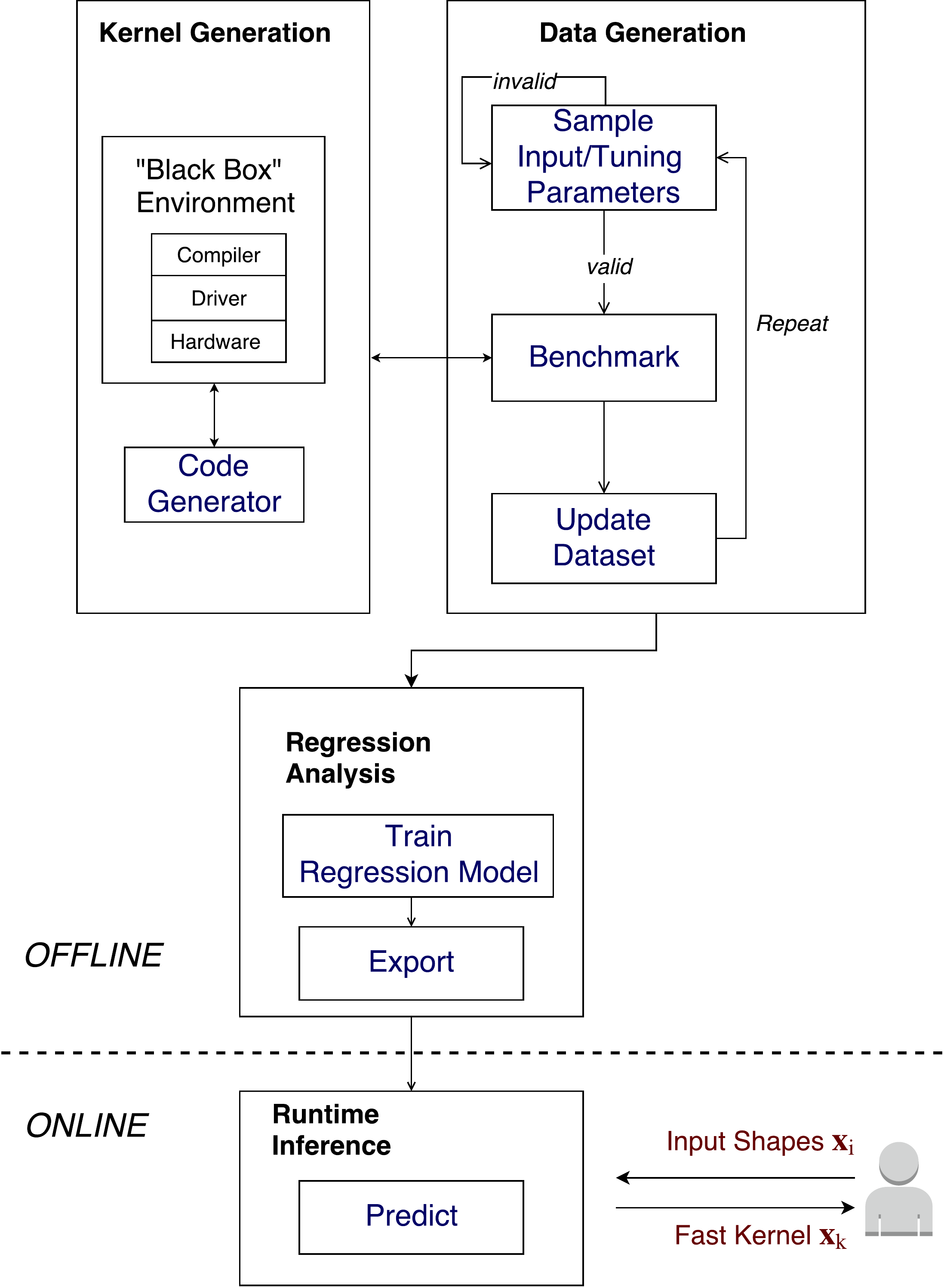}
\caption{Overview of ISAAC}
\label{fig:flowchart}
\end{figure}

This paper aims at offering a new perspective on automatic performance tuning. We present a system, ISAAC, which does not produce a fixed set of tuning parameters \textit{per se}, but rather a function that maps input characteristics to such parameters. We show that this function can be automatically learned from empirical benchmarking data using standard machine learning techniques (i.e., multi-layer perceptron), and propose a simple statistical method to speed-up the synthesis of a proper training dataset. An important addition of our framework is the use of a relatively low-level intermediate language (i.e., NVIDIA PTX), as opposed to higher-level alternatives typically used in similar systems (i.e., C, CUDA or OpenCL). While this is not strictly necessary and restricts our numerical experiments to NVIDIA GPUs only, this strategic choice leads to (1) better code generation, (2) faster compilation, and (3) more accurate performance models (due to simpler instruction selection heuristics).

Our system is composed of four major components (see Figure \ref{fig:flowchart}), each of which will be described in a separate section of this paper: Section \ref{sec:kernel-generation} describes the design and implementation of efficient code generation/parameterization techniques for matrix multiplication (GEMM) and convolution (CONV). Section \ref{sec:data-generation} defines the process by which we generate training data for the input-aware predictive model presented in Section \ref{sec:predictive-modeling}. Section \ref{sec:runtime-inference} shows how this model may be used at runtime to quickly infer globally optimal kernels given any input configuration. Section \ref{sec:numerical-benchmarks} provides a numerical evaluation of our system on various practical problems, and shows substantial performance gains over both cuBLAS and cuDNN (up to 3x), which we analyze in Section \ref{sec:analysis}. Finally, Section \ref{sec:conclusions} provides concluding remarks and directions for future work. Prior to diving into the details of our system, Section \ref{sec:related-works} provides insights on existing related work.

\section{Related Work} \label{sec:related-works}
Automatic performance tuning is a well established technique that has been effectively leveraged in a wide range of core HPC libraries (e.g., FFTW \cite{frigo05}, SPIRAL \cite{puschel04}, ATLAS \cite{whaley98} and OSKI \cite{vuduk05}). Despite their obvious merits, these projects largely focus on delivering portable performance across  architectures (ISA, memory hierarchies ...) rather than input properties (matrix sizes, sparsity patterns ...). This mismatch can result in the inefficient use of available hardware resources, ultimately leading to sub-optimal performance and/or energy efficiency.

Input-aware auto-tuning arose recently \cite{liu09} as a way to solve this issue, and has been since then applied to a variety of problems including poly-algorithmic selection \cite{ding15}, OpenACC loops optimization~\cite{magni13}, and general-purpose GPU compilers \cite{muralidharan14, samadi12}. This surge of interest is encouraging, but has yet to win over an industry dominated by manual heuristics. It is indeed common for high-budget vendor libraries (e.g., MKL, cuBLAS) to engineer a set of several highly-optimized assembly kernels, and handcraft heuristics for runtime kernel selection. In addition to being expensive and time-consuming, this process can create portions of the input-space where the performance is poorly optimized -- if at all. Our work, on the other hand, depicts a fully automated approach that not only fills such ``performance holes'', but also equals vendor libraries where they perform best (e.g., LINPACK).

\begin{figure}[h!]
\centering
\includegraphics[scale=.67]{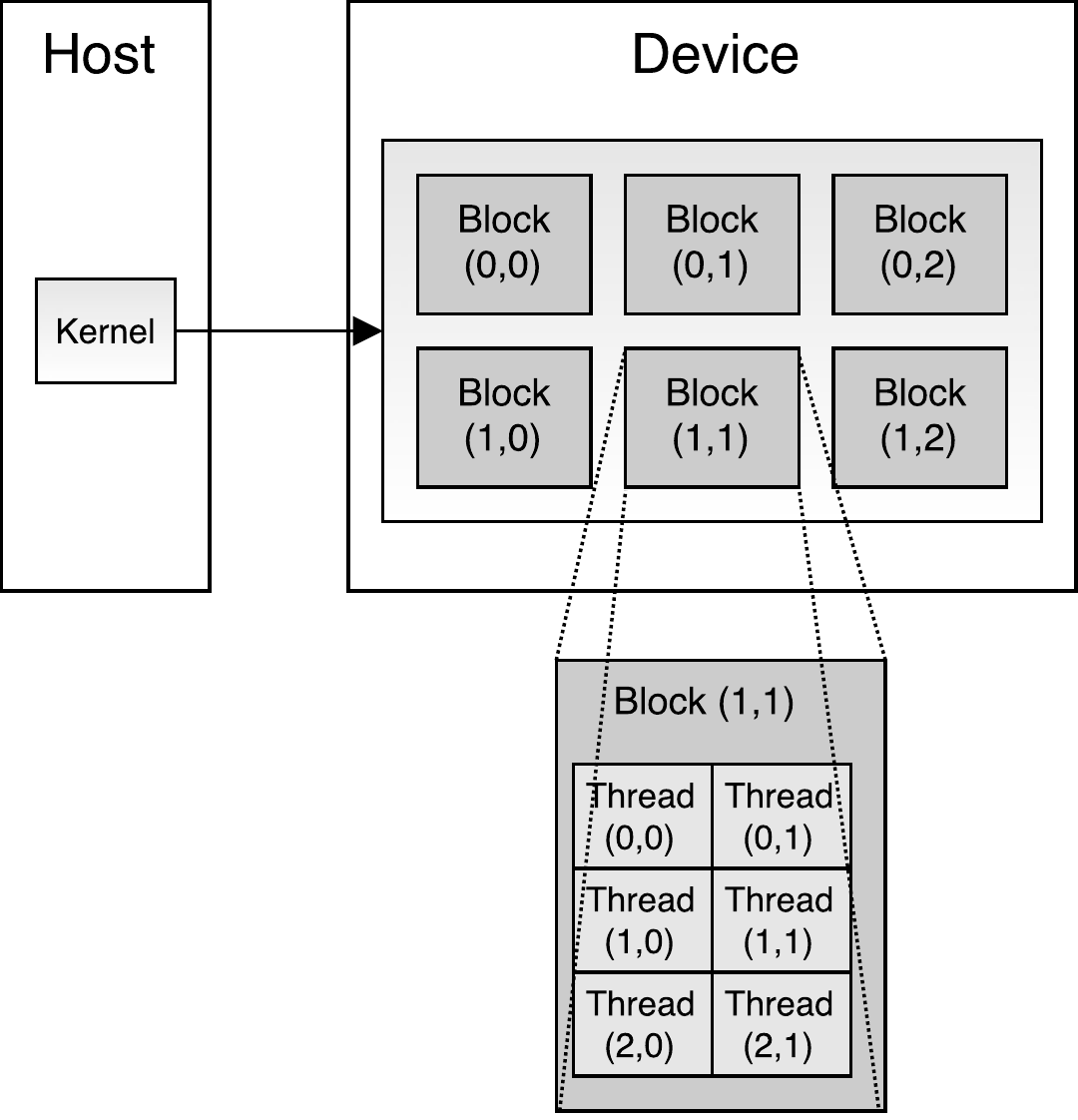}
\caption{Programming model for our kernel generator}
\label{fig:prog-model}
\end{figure}
\vspace{.5cm}

\section{Kernel Generation} \label{sec:kernel-generation}

In this section, we present the programming model underlying our forthcoming analysis, as well as the design and implementation of flexible source code parameterization techniques for GEMM and CONV. We introduce a set of reduction splitting parameters that improves our system's performance for deep reductions that may arise in, for instance, covariance matrices computations. This technique -- which is commonly found in communication-avoiding distributed algorithms for GEMM \cite{3d-gemm} -- is to our knowledge too often overlooked by automatically tuned on-node software libraries.

\subsection{Programming Model}
Fig.~\ref{fig:prog-model} shows the programming model assumed by our framework. This model has been adopted by many programming languages for multi/many-core devices, including PTX, CUDA and OpenCL.

Threads are arranged into a grid of 1-, 2- or 3-D blocks, where they may communicate using either synchronization barriers or shared memory. Communication between blocks is however only possible upon kernel completion through the use of global memory -- typically GDDR or HBM.

Each individual thread and block can be globally identified, thereby allowing different program instances to execute a given algorithm on different tiles of input data. The core idea behind auto-tuning frameworks is to parameterize the shape of these tiles, hence varying the amount of computation and resources used by each thread/block, ultimately fitting the underlying memory hierarchy and hardware threading mechanisms.

\subsection{Matrix Multiplication}
We now describe an input- and hardware- portable kernel parameterization for the matrix multiplication problem:
$$C = AB \qquad C\in \mathbb{C}^{M \times N}, ~A\in \mathbb{C}^{M \times K}, ~B\in \mathbb{C}^{K \times N}$$

Since this algorithm can be compute-bound for certain values of $(M, N, K)$, achieving high-performance requires efficient data-reuse and latency hiding. The former can be obtained via tiling and prefetching, while the latter necessitates thread level parallelism (TLP) and/or instruction level parallelism (ILP). 

For GPUs, TLP is implemented in hardware using a runtime scheduler: whenever a thread is stalled due to e.g., unfinished data transfers, another thread takes over the compute resources and execute another independent stream of instructions. This process actually happens at a granularity of 32 threads (i.e., a ``warp'') for NVIDIA hardware.

On the other hand, ILP is mostly handled in software. For the sake of energy efficiency, modern accelerators indeed outsource dependencies analysis to their respective Instruction Set Architectures (ISA): assembly programs are now often required to specify stall counts along with op-codes and operands.
\\
\begin{figure*}[t]
\centering
\includegraphics[scale=.6]{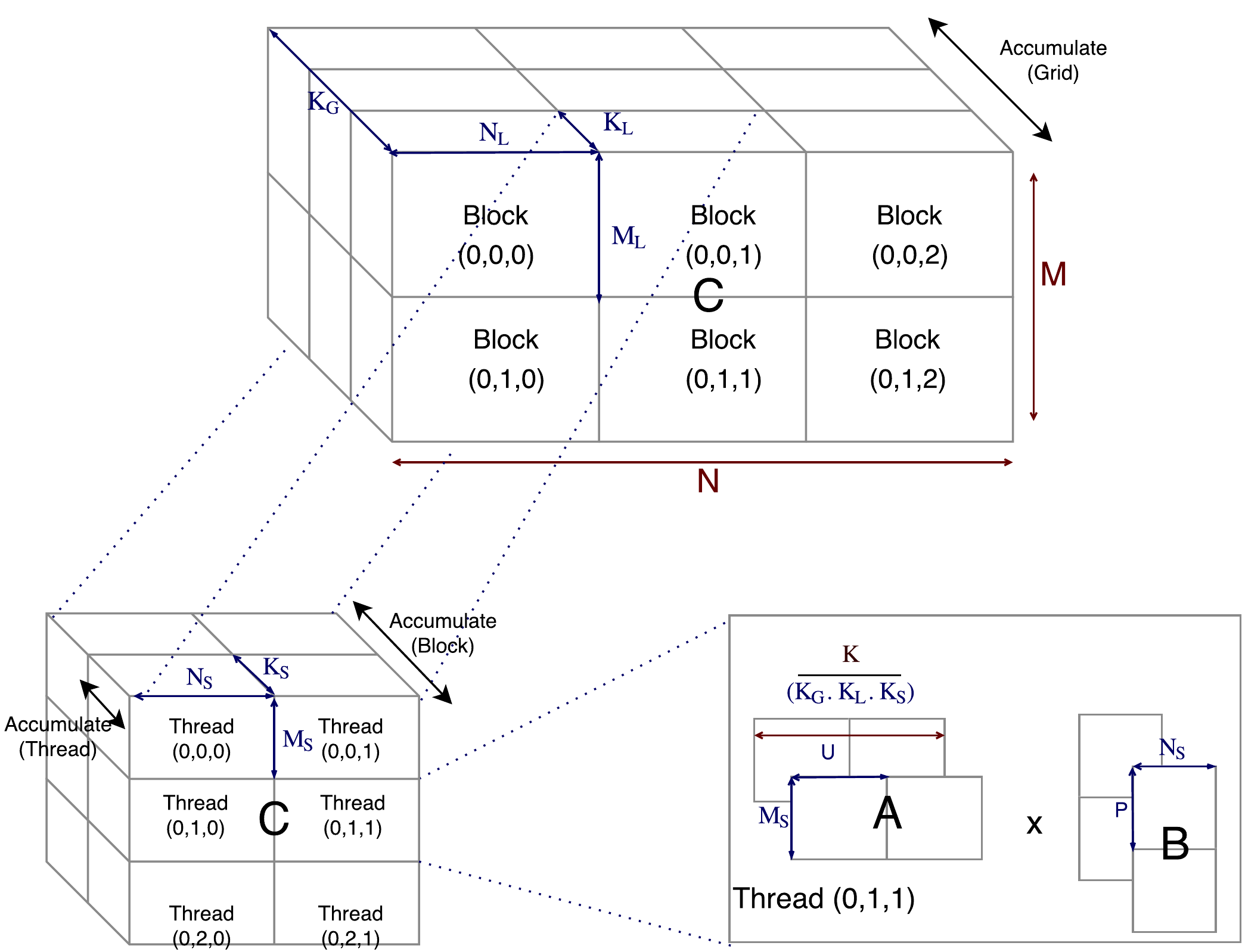}
\captionsetup{justification=centering}
\caption{Flexible parameterization of matrix multiplication. \\Input and tuning parameters are respectively shown in red and blue.}
\label{fig:gemm-template}
\end{figure*}

It is crucial to note that all these optimization techniques exhibit trade-offs with one another. Large tile sizes, for instance, promote data-reuse but require more hardware resources, potentially undermining TLP. On the other hand, if tiles are too small, independent instructions will become rare and opportunities for ILP will be reduced. This implies that, when the tiling factor along one direction is constrained to be ``small`` (due to the shape of the input matrices), it becomes necessary to mindfully increase tiling along another dimension to compensate. What it means for a tile to be ``small`` or ``large``, however, is a hidden property of the underlying hardware -- that even experts rarely fully understand. It should be clear, now, that optimal tile sizes depend not only on the target micro-architecture but also on user-provided input parameters not necessarily known in advance.

Figure \ref{fig:gemm-template} describes an algorithmic parameterization able to adjust these factors over a wide range of potential hardware architectures and input matrices. Each thread (resp. block) computes a tile of $M_S \times N_S$ (resp. $M_L \times N_L$) elements of $C$. In order maximize data-reuse, each work-group prefetches, into shared memory, $M_L \times U$ elements from $A$ and  $U \times N_L$ elements from $B$. These tiles can be transposed in-place if necessary. The actual computations are then performed in a fully unrolled fashion -- thereby producing a $K/U$ \emph{dependent} stream of $M_S . N_S . U$ multiply-accumulate instructions each.  Because $M_S$ and/or $N_S$ may be constrained to be very small in practice, it can become necesarry to create additional independent arithmetic instructions by splitting the computations along the reduction axis $K$, and accumulate the resulting partial results in a separate step. We therefore introduce three parameters $K_S$, $K_L$ and $K_G$ to split the reductions within respectively each thread, block and grid. Accumulation may then be performed using either registers addition, shared reductions or global atomics.

A common concern for practitioners is the handling of cases where $M$ (or $N$) is not a multiples of $M_L$ (or $N_L$). Fortunately, the use of predicated instructions in PTX makes it possible to perform bounds checking much more efficiently than with input padding. We will come back to this later.
\subsection{Multi-Channel Convolution}
We now consider the following convolution algorithm:
\begin{equation}
O_{k, :, :, n} = \sum_{c=0}^C I_{c, :, :, n} \star F_{c, :, :, k}
\label{eq:convolution}
\end{equation}
Where
\begin{align*}
O&\in R^{K \times P \times Q \times N}\\
I&\in R^{C \times H \times W \times N}\\
F&\in R^{C \times R \times S \times K}
\end{align*}

This operation  is a useful generalization of the usual 2D convolution operator $\star$: instead of convolving a single image with a single filter, it convolves a set of $C$ different images with $C$ different filters and and returns the sum of all the resulting matrices. This process is repeated for $N$ sets of images and $K$ sets of filters.

Due to the rise of Deep Learning over the past few years, this algorithm has become a bottleneck in many industrial and academic applications -- each operating on its own specific input domain. This has created a strong demand for input-aware peak performance that even the most popular frameworks fail to fully satisfy. It is indeed common for convolution libraries (e.g., cuDNN) to provide relatively poor performance on signal processing applications where (\ref{eq:convolution}) degenerates to $\star$ -- that is $N = C = K = 1$  -- or even certain standard benchmarks (e.g., DeepBench).

Since every element $O_{k,p,q,n}$ of the resulting tensor is the inner product of $CRS$ element from $I$ and $F$, it is possible to reformulate multi-channel convolutions as implicit matrix-multiplication problems: tiles loaded from $I$ and $F$ are scrambled while being stored to shared memory, using an indirection table in order to alleviate integer arithmetics in the algorithm's inner loop. 

It follows that we can use a parameterization similar to that exposed in Figure \ref{fig:gemm-template}, except that tiling is performed across five dimensions (K, P, Q, N, C) rather than three. Each thread (resp. block) computes a tile of $K_S \times P_S \times Q_S \times N_S$ (resp. $K_L \times P_L \times Q_L \times N_L$) elements of $O$. For the sake of data-reuse, each work-group prefetches, into shared memory, $N_L \times P_L \times Q_L \times U$ elements from $I$ and  $U \times K_L$ elements from $F$. The offsets for these load operations are obtained using the aforementioned indirection table. The actual computations are then performed in a fully unrolled fashion, and the reduction along $C$ is split using three tunable parameters: $C_S, C_L$ and $C_G$.

\section{Data Generation} \label{sec:data-generation}
Let $\mathbb{X}$ and $\hat{\mathbb{X}}$ be respectively the space of \emph{legal} and \emph{possible} configuration for the aforementioned parameterization schemes. This distinction is necessary because some kernels can be properly compiled but not safely executed  on the target device, due to the excessive usage of hardware resources such as shared memory or registers. For the GEMM algorithm described above, there are 10 tuning parameters and 6 input parameters -- 3 shapes, 1 data-type and 2 transposition layouts -- so $\mathbb{X} \subset \hat{\mathbb{X}} \subset \mathbb{N}^{16}$.

Input-aware auto-tuning works by building a device-specific regression model $\mathcal{R}$ for the performance of any combination of legal input and tuning parameters $\vx \in \mathbb{X}$. At runtime, the set of input parameters is fixed by the user, and $\mathcal{R}$ can be optimized over tuning parameters only. 

While $\mathcal{R}$ could technically be analytically approximated using deep expert knowledge, doing so could reduce its portability -- let alone performance portability -- across alternative and future micro-architectures. In this paper, we propose to learn $\mathcal{R}$ automatically from a large amount of benchmarking data obtained via the following statistical process.

\subsection{Generative Modeling}

Formally speaking, the goal of the data generation step is to produce a set of pairs $(\vx, y)$, where $\vx \in \mathbb{X}$, and $y \in \mathbb{R}$ is a performance measurement (e.g., FLOPS, Joules, FLOPS/W...) of the kernel induced by $\vx$ on the target hardware. When the number of parameters is small enough and the underlying resource constraints are known in advance, $\mathbb{X}$ can be pre-computed, in which case random parameter values can be trivially obtained via uniform sampling.

On the other hand, when only $\hat{\mathbb{X}}$ is explicitly known, uniform sampling can be extremely wasteful (For GEMM, more than 99.9\% of the resulting samples are illegal). A more tractable solution is to build a generative model $\mathcal{G}$ able to sample directly from the latent space of legal configurations $\mathbb{X}$.  It is easy to imagine scenarios where $\mathcal{G}$ would be defined by a complex graphical model, but this would require a thorough analysis that is beyond the scope of this paper. Instead, our framework uses a naive technique which is simple to understand yet significantly more efficient than uniform sampling.

Our generative model treats $\vx$ as a random vector whose components $\vx_i$ are independent categorical  variables. In other words, we assume: 
$$p(\vx \in \mathbb{X}) = p(\vx_0)p(\vx_1)\cdots p(\vx_N)$$

The probability distribution of each parameter $\vx_i$ can be approximated empirically, as the proportion of accepted values after a short period of uniform sampling. For instance, assuming that $\vx_1 = M_S$ may only take four values -- say, $1, 2, 4, 8$ -- which respectively appear 5, 20, 25 and 50 times out of 100 uniformly sampled valid configurations, our framework assigns:
\begin{align*}
p(\vx_1 = 1) = .05 \qquad p(\vx_1 = 2) = .2 \\
p(\vx_1 = 4) = .25 \qquad p(\vx_1 = 8) = .5
\end{align*}
Because we never really want any such probability to be exactly zero, we initialize each such count at a value $\alpha > 0$ (our implementation uses $\alpha = 100$). Formally speaking, this corresponds to assuming a Dirichlet prior distribution on $\vx_i$.

\subsection{Performance}
Table \ref{tab:sampling-performance} shows the proportion of invalid configuration generated by the above sampling method as compared to naive, uniform sampling, when each parameter is constrained to be a power of two between 1 and 16.

\begin{table}[h!]
\centering
\begin{tabular}{|c|c|c|}
\hline
& Categorical & Uniform\\
\hline
GEMM & 20\% & 0.1\% \\
\hline
CONV & 15\% & 0.1\% \\
\hline
\end{tabular}
\caption{Proportion of samples accepted by our categorical generative model vs uniform sampling}
\label{tab:sampling-performance}
\end{table}
We see that this model, although simplistic, offers large performance improvements over uniform sampling, reducing the amount of bad samples by more than two orders of magnitude. Using this method, we were able to benchmark 50,000 valid different kernels in less than two hours.

Again, we emphasize that this model is only meaningful when $\mathbb{X}$ cannot be pre-computed.

\begin{figure*}[tp]
\centering
\includegraphics[scale=.4]{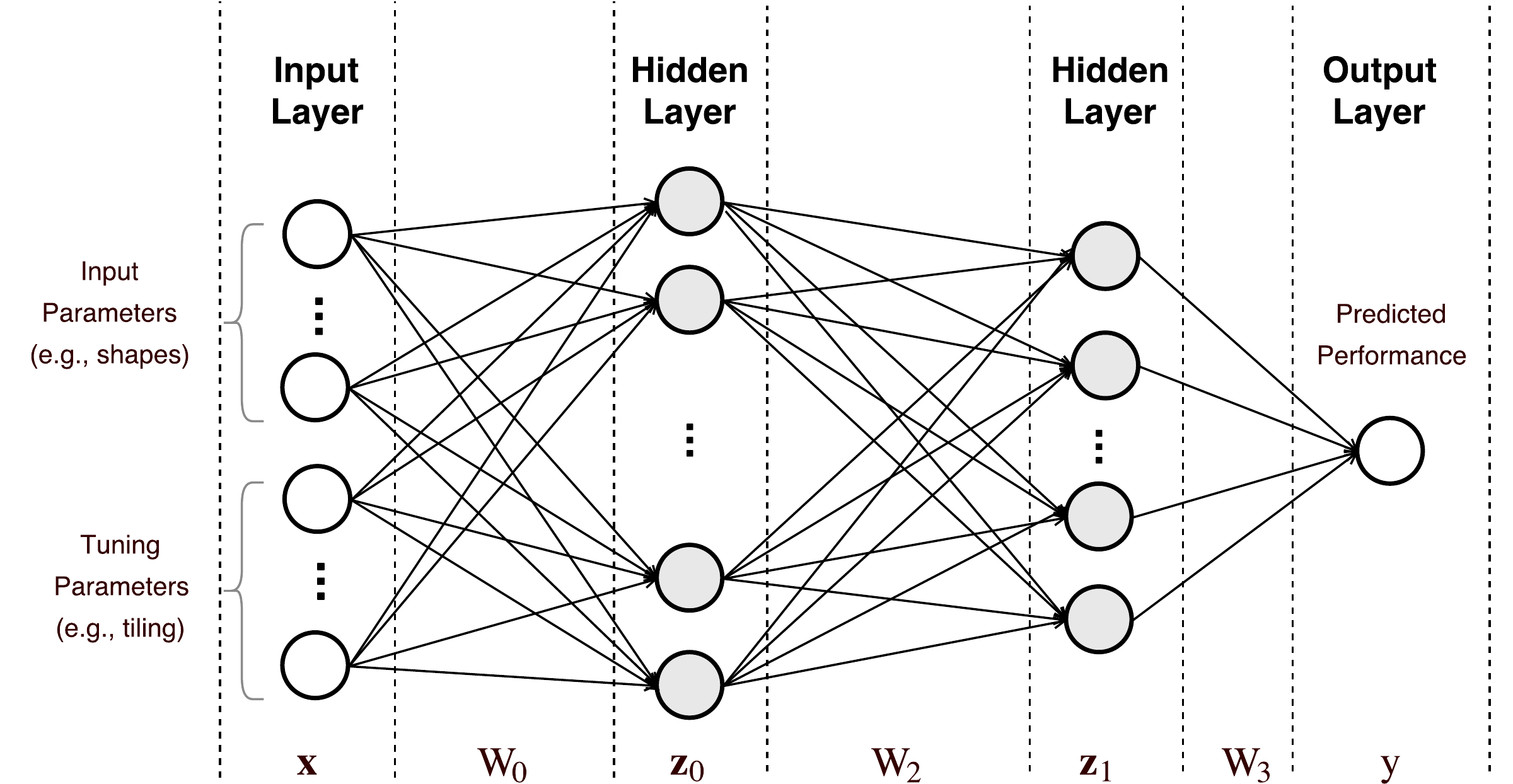}
\caption{A Multi-Layer Perceptron}
\label{fig:mlp}
\end{figure*}

\section{Regression Analysis} \label{sec:predictive-modeling}
Once a sufficient amount of training data $\mathcal{X}$ has been gathered using the above sampling method, our system builds a predictive model for the performance of any parameter vector $\vx \in \mathbb{X}$. This is known as \textit{regression analysis}.

We evaluated multiple potential solutions before opting for a multi-layer perceptron (MLP), as it (1) scales best with large datasets (given enough time and resources, our dataset can be made arbitrarily large) and (2) naturally handles common nonlinearities found in performance modeling such as maximums and minimums. 

Furthermore, since MLP involving small feature vectors (around 20 in our case) rely on highly rectangular matrix computations, our system could itself be bootstrapped to make its own auto-tuning procedure more efficient.

\subsection{Multi-Layer Perceptron}
Fig.~\ref{fig:mlp} shows the architecture of a basic MLP. The mapping from $\vx$ to $y$ is organized in multiple layers of nonlinearly-activating nodes. Successive layers are fully connected, meaning that each node $i$ in one layer $L_n$ connects to each node $j$ in the following layer $L_{n+1}$ with a trainable weight $(W_n)_{i,j}$.

In other words, $y$ can be computed from $\vx$ using the following algorithm:

\begin{algorithm}
\KwIn{Weights $W_0, W_1, \cdots,W_{L-1}$ ; Features $\vx$.}
\KwOut{Performance prediction $\hat{y}$}    
$\va_{-1} = \vx$\;
\For{$n\leftarrow 0$ \KwTo $L-1$}{
	$\vz_n = W_n \va_{n-1}$\;
	$\va_n = f_n(\vz_n)$
}
$\hat{y} = \va_{L-1}$
\caption{Forward Propagation}
\end{algorithm}
Where $f_i$ is a non-linear activation shared by all the neurons in layer $n$. We emphasize that, under this model, multiplicative relationships between different elements of $a_n$ cannot be easily modeled. We will come back to this later.

The parameters $W_i$ are chosen so as to minimize a given loss $\sum_\vx \mathcal{L}(\hat{y}(\vx), y(\vx))$ on the predicted output. For regression analysis, it is desired that the predictions $\hat{y}(\vx)$ be noisy estimates of the true outputs $y(\vx)$, leading to the mean square error (MSE) loss function (when the noise is Gaussian). Since $\mathcal{L}$ is always chosen to be differentiable, this minimization can be carried out using e.g., Stochastic Gradient Descent.

\subsection{Implementation details}
Before explaining the details of our MLP implementation (hyperparameters, feature transformation, non-linearities), it can be useful to review the existing literature about GPU performance modeling. A comprehensive review of analytical performance models for compute-bound GPU kernels was offered by Volkov in his doctoral dissertation \cite{volkov-thesis}. 

A common strategy for estimating the average arithmetic and memory throughput (in instructions/cycles) of the target kernel $K$ is:
\begin{align*}
t_\text{arith}(n) &= \max\Big(\frac{\text{alu\_latency}}{n}, \text{alu\_throughput}\Big) \\
t_\text{mem}(n) &= \max\Big(\frac{\text{mem\_latency}}{n}, \text{mem\_throughput}\Big) \numberthis \label{eq:model-1}
\end{align*}

Where $n$ is the mean occupancy (in warps per multi-processor), and $\text{alu\_throughput}$, $\text{mem\_throughput}$ are underlying hardware characteristics.

The total execution time $t(n)$ of $K$ is then:
\begin{equation}
t(n) = \max(t_\text{arith}(n)i_\text{arith}, \quad t_\text{mem}(n)i_\text{mem})
\label{eq:model-2}
\end{equation}
Where $i_\text{arith}$ and $i_\text{mem}$ are respectively the number of arithmetic and memory instructions in K.

The entire premise of our approach is that all the quantities involved in these computations depend -- more or less strongly -- on the relationship between \emph{hidden} hardware features (e.g., number of ALU, memory bandwidth, maximum throughput, banking structures) and \emph{known} input/tuning parameters (e.g., tensor shapes, tile sizes).  A successful MLP should (implicitly) learn not only these relationships but also the corresponding hidden variables.

As suggested by (\ref{eq:model-1}) and (\ref{eq:model-2}), it is expected that the relationships between all these variables include multiplications, divisions and maximums. Because, as mentioned previously, neural networks are not naturally designed to handle multiplications between different features,  setting $\va_{-1} = \log(\vx)$ greatly improved the performance of our system. Furthermore, choosing the rectified linear unit (relu)  activation function $f_i(\vz_i) = \max(0, \vz_i)$ seems appropriate to handle maximums.

It may seem at first sight that deeper and wider MLPs would lead to higher runtime latency. However, research on neural networks inference tends to show that it is preferrable to train larger networks even if it means pruning or binarizing them afterwards \cite{binarized-nets}.

\subsection{Accuracy}
A common criticism of neural networks is that they are hard to engineer, hence this section attempts to provide insights on how good MLP architectures may be designed for our problem, as well as intuition regarding the amount of training data necessary to achieve good performance. We used matrix multiplication for our analysis, but the same qualitative behavior was observed in convolutions.

Table \ref{tab:cross-validation-mse} shows the cross-validation MSE of several MLP architectures, as measured on a fixed set of $10,000$ data-points separate from the $200,000$ samples used for training. Unsurprisingly, deeper networks seem to perform much better than shallower one (given a fixed amount of parameters). The accuracy of the network can be adjusted by adding (moderately wider) layers, at the cost of longer training and higher runtime latency. We emphasize the importance of the logarithmic feature transformation exposed in the previous subsection, without which our system would converge to much worse solutions -- if at all.
\begin{table}[h!]
\centering
\begin{tabular}{|c|c|c|c}
\hline
Hidden layer sizes&\#weights & MSE (no log)\\
\hline
64&1k& 0.17 (1.2)\\
512&10k&0.13 (1.0)\\
32, 64, 32 & 5k&0.088 (0.80)\\
64, 128, 64 & 17k&0.08 (0.75)\\
32, 64, 128, 64, 32 & 21k & 0.073 (-)\\
64, 128, 256, 128, 64 & 83k & 0.067 (-)\\
64, 128, 192, 256, 192, 128, 64 & 163k & 0.062 (-)\\
\hline
\end{tabular}
\captionsetup{justification=centering}
\caption{Cross-validation MSE for various MLP architectures}
\label{tab:cross-validation-mse}
\end{table}

\begin{figure}[h!]
\centering
\includegraphics[width=\columnwidth]{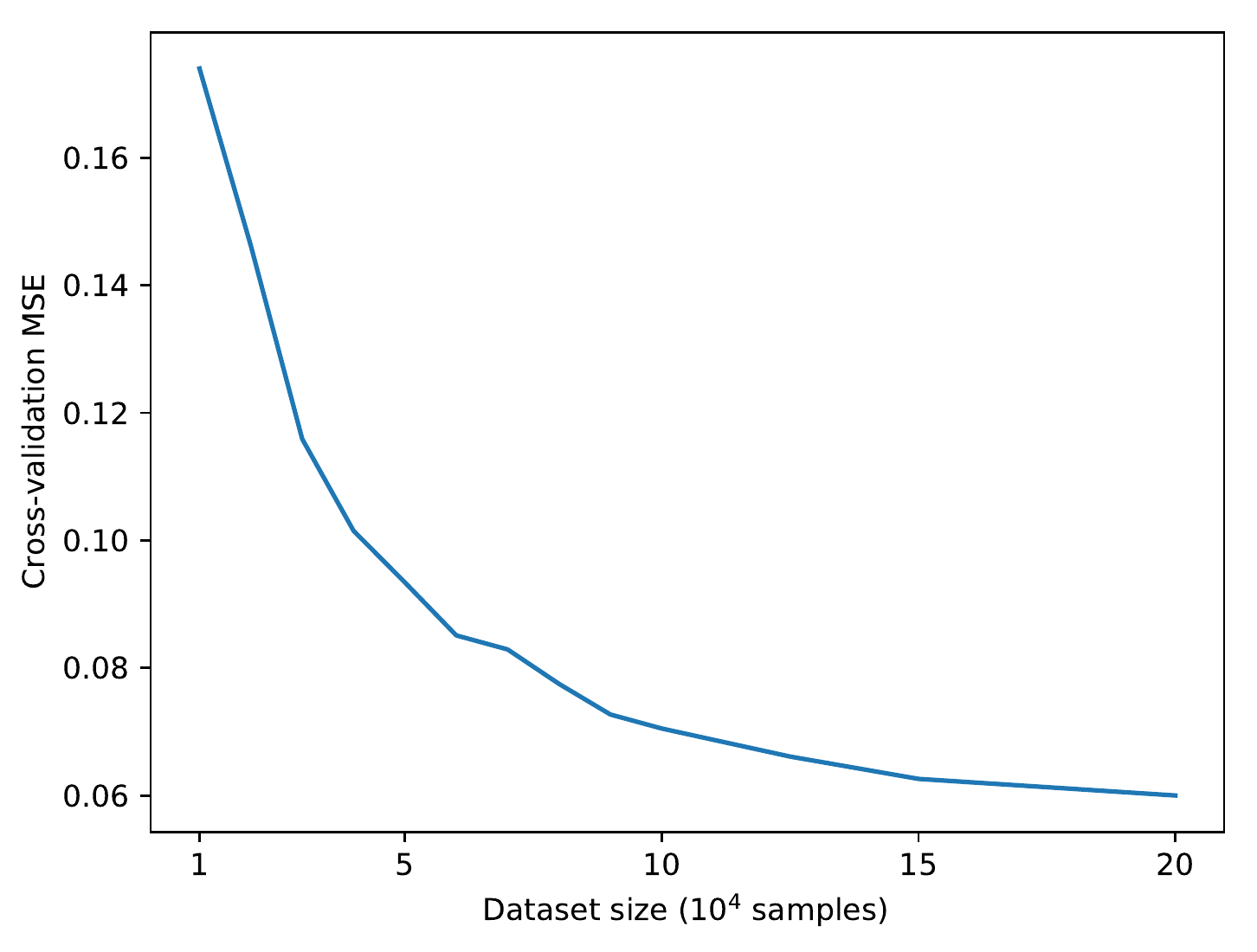}
\captionsetup{justification=centering}
\caption{Cross-validation MSE for various data-set sizes}
\label{fig:cross-validation-mse}
\end{figure}

Figure \ref{fig:cross-validation-mse} shows the evolution of our most accurate MLP's accuracy as the amount of training data available grows. As expected, collecting more data does not seem to provide much benefits beyond a certain point (150,000 samples for GEMM, or $\sim6$ hours of data collection).

 \section{Runtime Kernel Inference} \label{sec:runtime-inference}
At this point, we possess a trained regression model that can predict the performance of any combination of input and tuning parameters. This model can be evaluated very quickly, in parallel, and with constant latency. This differs from actual kernel executions on a GPU, which may be slow, lock the device or even time-out when very inefficient kernels meet large problems.

At runtime, the input parameters are provided by the user and fixed. Our model can be optimized over the remaining (i.e., tuning) parameters. Any discrete optimization method (e.g., simulated annealing, genetic algorithm, exhaustive search) may be used for this purpose. In this paper, we have opted for an exhaustive search, as it has several attractive properties when the number of tuning parameters is low enough:
\begin{itemize}
\item It is guaranteed to find the global optimum within the specified search range.
\item The search is highly parallelizable. Up to a million different configurations per second can be evaluated -- and potentially more, shall our system be bootstrapped in the process.
\item It is trivial to obtain the 100 (or more) fastest configurations for our model, and re-evaluate them on the target GPU to smooth out the inherent noise of our predictive model.
\end{itemize}
The cost of exhaustive runtime inference, while high -- up to a few seconds -- is several orders of magnitude faster than running an exhaustive search on the target hardware (which can take up to 10 hours).

The resulting predictions may be used directly in applications where this latency would be negligible (e.g., Deep Learning), cached on the filesystem, or even used as a kernel generation backend for low-latency libraries such as cuBLAS or cuDNN.

\section{Numerical Experiments} \label{sec:numerical-benchmarks}
 
This section explores the performance of ISAAC for various input shapes, data-types and transposition layouts  covering typical problem dimensions found in industrial benchmarks (LINPACK), scientific computing (LAPACK), deep learning (DeepBench) and signal processing (Independent Component Analysis).

\subsection{Hardware architectures}
While the focus of this paper is set on input-awareness, it is important that our framework be performance-portable across existing and future micro-architectures, hence our numerical experiments will be repeated on two distinct GPUs:

\begin{table}[h]
\centering
\resizebox{\columnwidth}{!}{
\begin{tabular}{ | l | c | c | r |}
  \hline
  & \bf Maxwell &  \bf Pascal \\
  \hline
  \bf GPU & GTX 980 TI & Tesla P100 (PCIE)\\
  \bf Market Segment & Consumer &  Server\\
  \hline
  \bf Micro-architecture & GM200 & GP100\\
  \bf CUDA cores & 2816 & 3584 \\
  \bf Boost frequency & 1075 MHz & 1353 MHZ \\
  \bf Processing Power & 5.8 TFLOPS & 9.7 TFLOPS \\
  \hline
  \bf Memory quantity & 6 GB & 16 GB \\
  \bf Memory Type & GDDR5 & HBM2 \\
  \bf Memory Bandwidth & 336 GB/S & 732 GB/s \\
  \bf TDP & 250W & 250W\\
  \hline
\end{tabular}
}
\caption{Test platforms hardware}
\label{tab:test-hardware}
\end{table}
 
These two devices, though both designed by NVIDIA within a span of two years, differ in many ways. First, the Tesla P100 offers much more processing power than the GTX980 TI, as its higher power-efficiency allows it to carry more CUDA cores running at a higher frequency. Second, the P100 offers two times the bandwidth of the GTX980 TI -- and again, these gains stem from major technological improvements. It is worth pointing out that HBM2 (large bus width, low frequency) and GDDR5 (small bus width, high frequency) handle memory transfers in a radically different way, to the point where IO-bound code designed for GDDR5 is not guaranteed to perform well with HBM2.


\subsection{Experimental protocol}
We compare our framework against cuBLAS 8.0 and cuDNN v6.0, which are the latest versions available at the time this paper is written. Despite a lot of research in automatic performance tuning, these two libraries have remained the gold standard for Linear Algebra and Deep Learning. Both libraries rely on handcrafted heuristics for choosing among a set of statically optimized assembly implementations.

The cuBLAS API exposes functionalities to manually call individual kernels via the \lstinline|cublasGemmEx| function, effectively allowing us to bypass any existing heuristics. We use this feature (under the label ``Best Kernel``) to discriminate bad heuristical choices from missing tiling schemes. 

We use the flag \lstinline|IMPLICIT_PRECOMP_GEMM| to force cuDNN to use of the algorithm presented in Section \ref{sec:kernel-generation}, with a scratch space of 64MB that remains on the device throughout the entire duration of our benchmark.

\subsection{GEMM Performance}
General Matrix Multiplication (GEMM) sits at the heart of High-Performance Computing, and is crucial to many applications, including supercomputer performance assessment, machine learning, signal processing and scientific computing. In this section, we evaluate our proposed framework on a set of input configurations (see Table \ref{tab:gemm-configurations}) that we believe are representative of its practical usage.
\begin{table}[h!]
\centering
\resizebox{\columnwidth}{!}{
\begin{tabular}{| l | l | l | l | l | l |}
\hline
\bf \bf M & \bf N & \bf K & \bf A-T & \bf B-T & \bf Description  \\
\hline
\multicolumn{6}{|c|}{LINPACK}\\
\hline
512 & 512 & 512 & No & Yes & Square case\\
1024 & 1024 & 1024 & No & Yes & Square case\\
2048 & 2048 & 2048 & No & Yes & Square case\\
\hline
\multicolumn{6}{|c|}{DeepBench}\\
\hline
2560 & 16 & 2560 & \{No, Yes\} & No & \shortstack{\{Forw/Back\} Propagation}\\
2560 & 32 & 2560 & \{No, Yes\} & No & \shortstack{\{Forw/Back\} Propagation}\\
2560 & 64 & 2560 & \{No, Yes\} & No & \shortstack{\{Forw/Back\} Propagation}\\
2560 & 128 & 2560 & \{No, Yes\} & No & \shortstack{\{Forw/Back\} Propagation}\\
\hline
\multicolumn{6}{|c|}{Independent component analysis (ICA) }\\
\hline
32 & 32 & 60000 & No & Yes & 32-channels\\
64 & 64 & 60000 & No & Yes & 64-channels\\
256 & 256 & 60000 & No & Yes & 256-channels\\
\hline
\multicolumn{6}{|c|}{\shortstack{LAPACK (Blocked SVD -- block-size 32 \cite{lahabar09})}}\\
\hline
4096 & 4096 & 32 & No & Yes & Iteration 0\\
3456 & 3456 & 32 & No & Yes & Iteration 64\\
896 & 896 & 32 & No & Yes & Iteration 100\\
\hline
\end{tabular}
}
\caption{Tasks considered for the evaluation of ISAAC on GEMM. The column 'A-T' (resp. 'B-T') is marked as 'Yes' if $A$ (resp. $B$) is transposed, and 'No' otherwise.}
\label{tab:gemm-configurations}
\end{table}

\subsubsection{GTX 980 TI}
\begin{figure}[h!]
	\includegraphics[width=.5\textwidth]{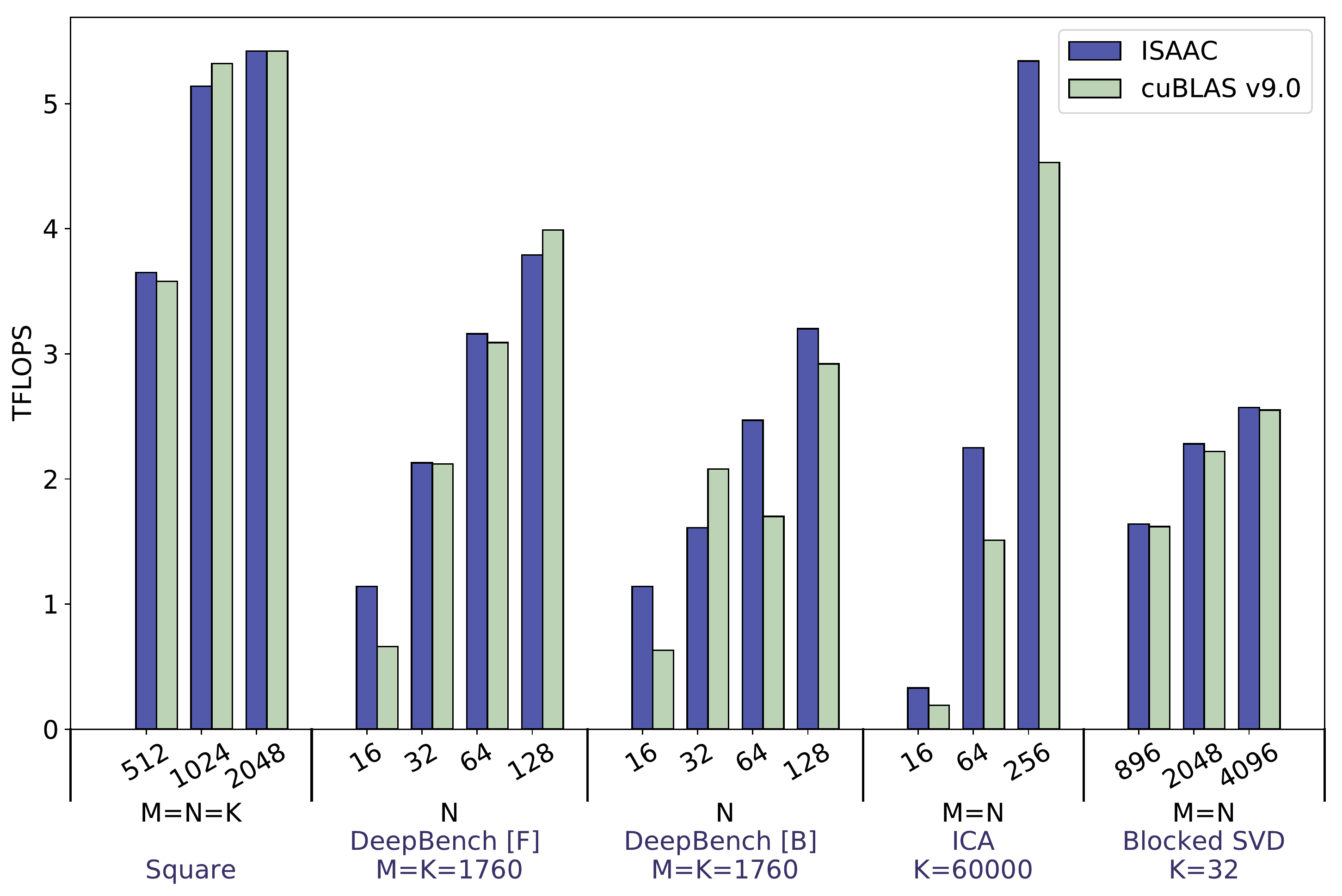}
	\caption{SGEMM performance on the GTX 980 TI}
	\label{fig:bench:gemm:980ti}
\end{figure}

The results of our benchmarks are shown in Figure \ref{fig:bench:gemm:980ti}.

\textbf{LINPACK}\\
Our system rivals cuBLAS's assembly kernels for large, square matrices (a case the library is specifically optimized for, due to its importance in performance assesment) and even outperforms it by almost 25\%  when M=N=K=512.

\textbf{DeepBench (Forward)}\\
The benefits of input-aware auto-tuning become more apparent for problems involving irregular input shapes. Our benchmark shows $80\%$ speed-ups on DeepBench for $N=16$ (here we show $M=N=2560$, but our results hold as long as $M, N$ are big enough to make GPU execution meaningful). These gains vanish as the batch size approaches tiling factors provided explicitly by cuBLAS ($N_L \in \{64, 128\}$). It should be nonetheless noted that large batch sizes are rarely used in practice due to bad convergence properties \cite{tang2017}. 

We note poor heuristical kernel selection for cuBLAS when $N \in \{32, 64\}$. Further investigation revealed that it was due to poor handling of reduction-splitting in the library's heuristics.

\textbf{DeepBench (Backward)}\\
We found reduction splitting ($K_L > 1$, $K_G > 1$) to be even more necessary for achieving good performance on DeepBench's back-propagation problems. This is due to unfavorable access patterns which requires both A and B to be internally transposed in shared memory prior to any computation. The latency of these transpositions can be hidden by using more warps, which is the exact purpose of reduction-splitting. All things considered, our framework outperforms cuBLAS's best kernel by $65\%$ when $N=16$ and by $35\%$ when $N=128$.

\textbf{ICA}\\
It is known that cuBLAS implements some form of global reduction splitting ($K_G > 1$) to handle cases where $K$ is large and $M.N$ is small. There seems to be several instances in which the library's heuristics fail to properly leverage this feature, resulting in drastic slow-downs (over an order of magnitude) in our ICA benchmarks. Even after bypassing kernel selection, cuBLAS remains $10\%$ slower than ISAAC, which is attributed to cuBLAS not implementing reduction splitting within streaming multi-processors ($K_L > 1$).

\textbf{LAPACK}\\
Minor performance gains (10\%) are observed for packed outer-products commonly found in blocked linear algebra algorithms (e.g., householder bi-diagonalization in SVD).

\begin{figure}[h!]
	\includegraphics[width=.5\textwidth]{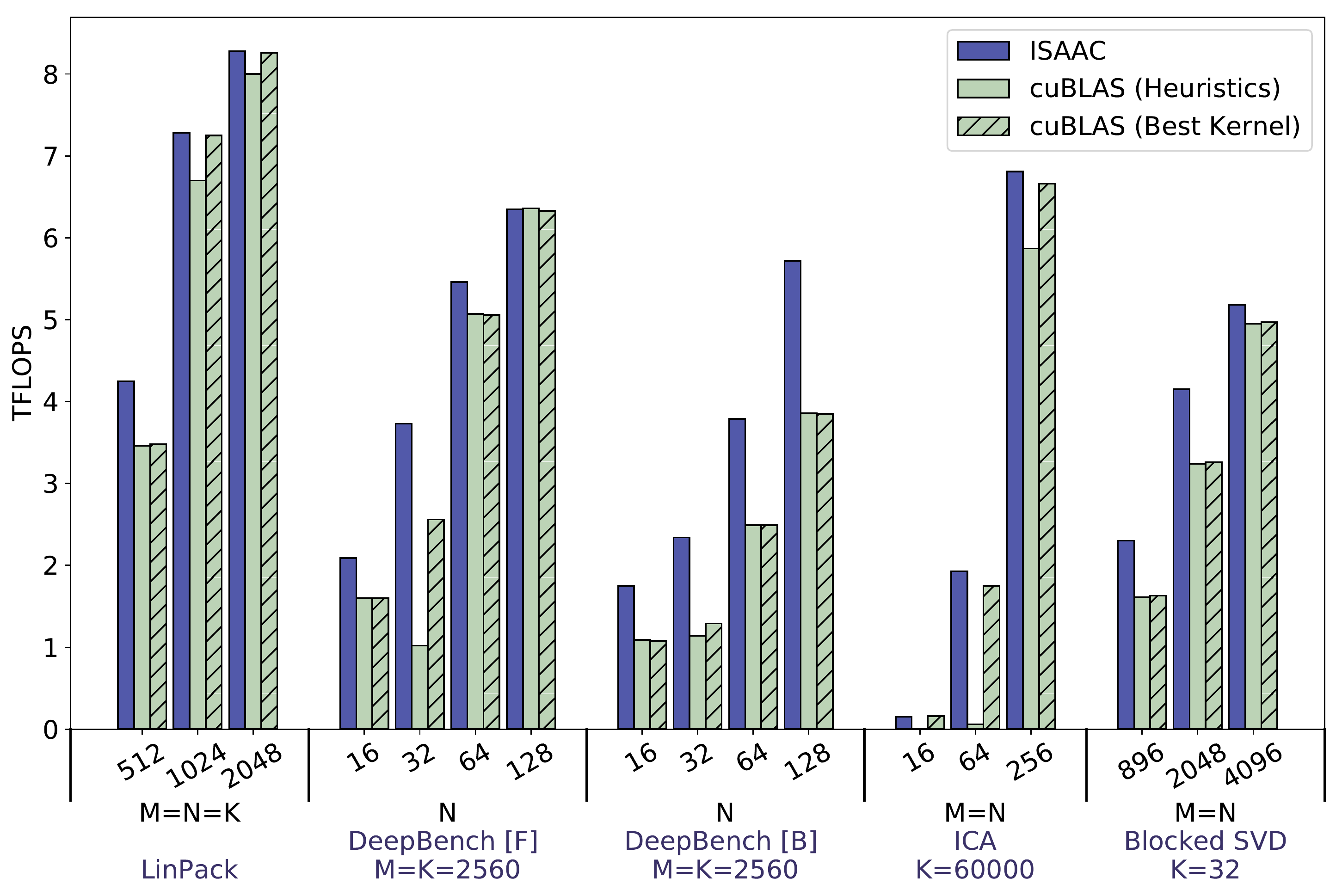}
	\caption{SGEMM performance on the Tesla P100}
	\label{fig:bench:gemv:amd}
\end{figure}

\subsubsection{Tesla P100}

\vspace{6pt}
\textbf{Single Precision}\\
Figure \ref{fig:bench:gemm:980ti} showed that cuBLAS achieves more than 90\% of Maxwell's peak performance on large, square matrices. This efficiency does not seem to carry over to Pascal, as cuBLAS saturates at 85\% of the P100's peak performance. On the other hand, our system's efficiency is constant ($85\%$), leading to performance parity with cuBLAS in our most pessimistic benchmarks.  The automation inherent to our approach also allows for shorter development cycles -- the tuning procedure only takes a few hours -- which could facilitate the deployment of software updates following the release of a new architecture.

The performance gains of ISAAC over cuBLAS's best kernel remain otherwise consistent with those observed in the previous subsection, reaching 25\% on LINPACK,  80\% on DeepBench, 5\% on ICA and 30\% on LAPACK. The heuristics used by cuBLAS seems to retained the same deficienies as for Maxwell.

\begin{figure}[h!]
\centering
\includegraphics[width=\columnwidth]{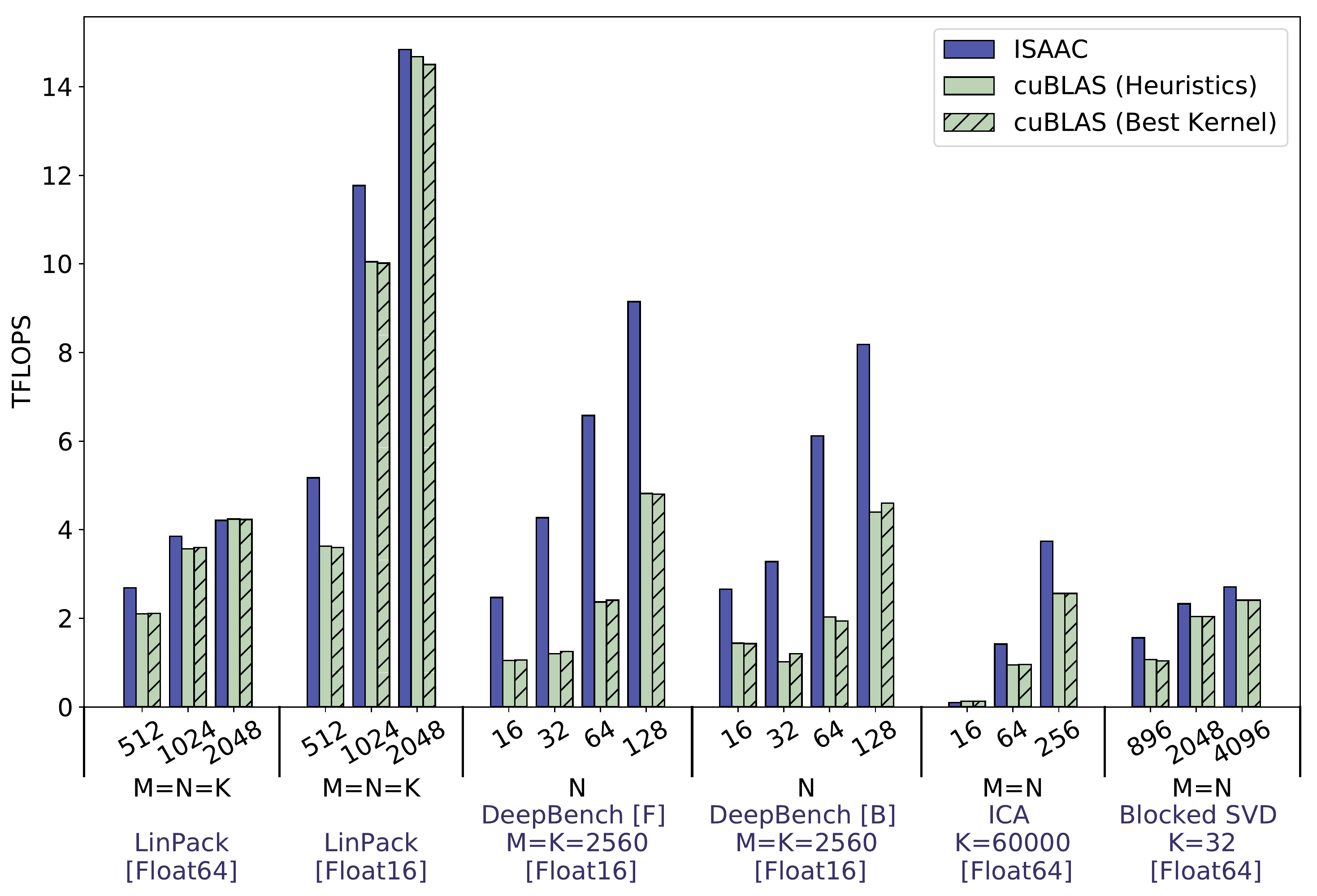}
\caption{H/DGEMM performance on the Tesla P100}
\label{fig:gemm-fp16-fp64}
\end{figure}

\textbf{Half/Double Precision}\\
Our numerical experiments would be incomplete without a proper account of ISAAC's half and double precision performance, as the usage of single precision arithmetics is discouraged in both Deep Learning (where half precision is sufficient) and Scientific Computing (where double precision is necessary). Fortunately, our approach is not bound to any particular data-type, hence we re-evaluate GEMM in half and double precision on the Tesla P100, which provides respectively 0.5x and 2x single precision peak performance for these cases. Half precision is used for DeepBench and LINPACK; double precision is used for the rest.

As shown in figure \ref{fig:gemm-fp16-fp64}, our framework retains significant performance gains over cuBLAS in double precision, averaging 5\% on LINPACK, 40\% on ICA and 15\% on LAPACK.

A major advantage of our framework is its ability to generate many different kernels at a very low cost. This inherent flexibility translates into tremendous performance gains in cases where adding support for new tiling schemes and/or specialized instructions is cumbersome, and apparently not implemented in cuBLAS. As a result, ISAAC is able to leverage the ``fp16x2`` instructions across the entire input-space, resulting in 2.5-3x speedups over cuBLAS on DeepBench. The near-optimal half-precision performance of NVIDIA's library on LINPACK underlines the existence of a limited set of NVIDIA kernels implementing this feature.

\subsection{CONV Performance}

The rise of Deep Learning over the last 5 years \cite{khrizhevsky2012} has made fast convolution routines not only desirable but also necessary to the rapid evolution of the field as a whole. CuDNN offers state-of-the-art performance for this algorithm, and is used in all major Deep Learning Frameworks (e.g., Tensorflow, Theano, Pytorch...).

In this section, we show that input-aware auto-tuning can be used to produce compute kernels sometimes faster than cuDNN. The network architectures considered in this section were extracted from the DeepBench suite so as to span 6 different concrete applications. 

The corresponding data shapes are shown in Table \ref{tab:conv-configurations}. Recall that cuDNN treats (N, P, Q, K, C, R, S) convolutions as implicit (NPQ, K, CRS) matrix multiplications.

\begin{table}[h!]
\centering
\resizebox{\columnwidth}{!}{
\begin{tabular}{| l | l | l | l | l | l | l | l| l| l|}
\hline
\bf N & \bf P & \bf Q & \bf K & \bf C & \bf R & \bf S & \bf NPQ & \bf CRS & \bf Name\\
\hline
\multicolumn{10}{|c|}{DeepSpeech}\\
\hline
16 & 79 & 341 & 32 & 1 & 5 & 20 & 431024 & 100 & Conv1\\
16 & 38 & 166 & 32 & 32 & 5 & 10 & 100928 & 1600&  Conv2\\
\hline
\multicolumn{10}{|c|}{OCR}\\
\hline
16 & 24 & 240 & 32 & 16 & 3 & 3 & 92160 & 144 & Conv3\\
16 & 12 & 120 & 64 & 32 & 3 & 3 & 23040 & 288 & Conv4\\
\hline
\multicolumn{10}{|c|}{Face Recognition}\\
\hline
8  & 54 & 54 & 64 & 64 & 3 & 3 & 23328 & 576 & Conv5\\ 
8 & 27 & 27 & 128 & 128 & 3 & 3 & 5832 & 1152 & Conv6\\ 
16 & 14 & 14 & 48 & 512 & 5 & 5 & 3136 & 12800 & Conv7\\
16 & 7 & 7 & 128 & 832 & 5 & 5 & 784 & 20800 & Conv8\\
\hline
\multicolumn{10}{|c|}{Vision}\\
\hline
8 & 112 & 112 & 128 & 64 & 3 & 3 & 100352 & 576 & Conv9\\ 
8 & 56 & 56 & 256 & 128 & 3 & 3 & 25088 & 1152 &  Conv10\\  
\hline
\multicolumn{10}{|c|}{Speaker ID}\\
\hline
16 & 128 & 39 & 174 & 64 & 5 & 5 & 79872 & 1600 & Conv11\\ 
16 & 256 & 19 & 87 & 128 & 5 & 5 & 77824 & 3200 &  Conv12\\
\hline
\multicolumn{10}{|c|}{ResNET}\\
\hline
16 & 7 & 7 & 512 & 512 & 3 & 3 & 784 & 4608 & Conv13\\
16 & 7 & 7 & 2048 & 1024 & 1 & 1 & 784 & 1024 & Conv14\\
\hline
\end{tabular}
}
\caption{Tasks considered for evaluating ISAAC on CONV}
\label{tab:conv-configurations}
\end{table}

\begin{figure}[h!]
	\includegraphics[width=.5\textwidth]{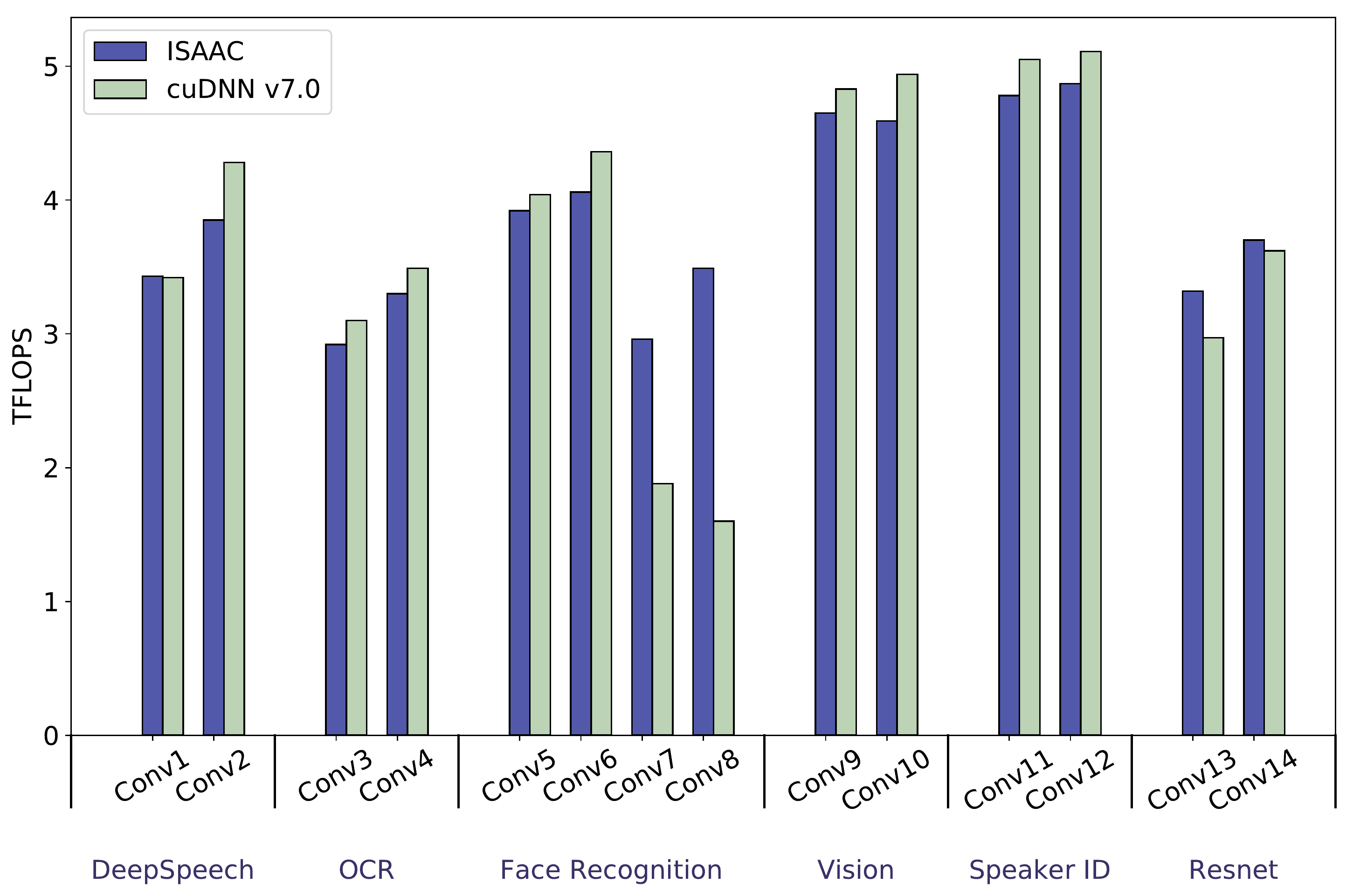}
	\caption{SCONV performance on the GTX 980 TI}
	\label{fig:bench:conv:gtx980}
\end{figure}

\subsubsection{GTX 980}
The performance benefits of ISAAC (see Figure \ref{fig:bench:conv:gtx980}) are noticeable but not as large as they were for GEMM. This is because cuDNN was optimized from the ground up with both Maxwell and DeepBench-like problems in mind (Large NPQ, small K and intermediate CRS).

 Nonetheless, we note substantial performance gains ($1.5\times$ to $2\times$) over cuDNN for the deep reductions  found in Conv7 and Conv8. Note that cuDNN provides no public way of benchmarking individual kernels, hence it is difficult to say whether these gains come from poor heuristical choices or missing tiling configurations.
 
 We also note appreciable speed-ups ($\sim10\%$) when NPQ is small and the operation does not degenerate to direct matrix multiplication ($RS > 1$, Conv13).

\subsubsection{Tesla P100}

\begin{figure}[h!]
	\includegraphics[width=.5\textwidth]{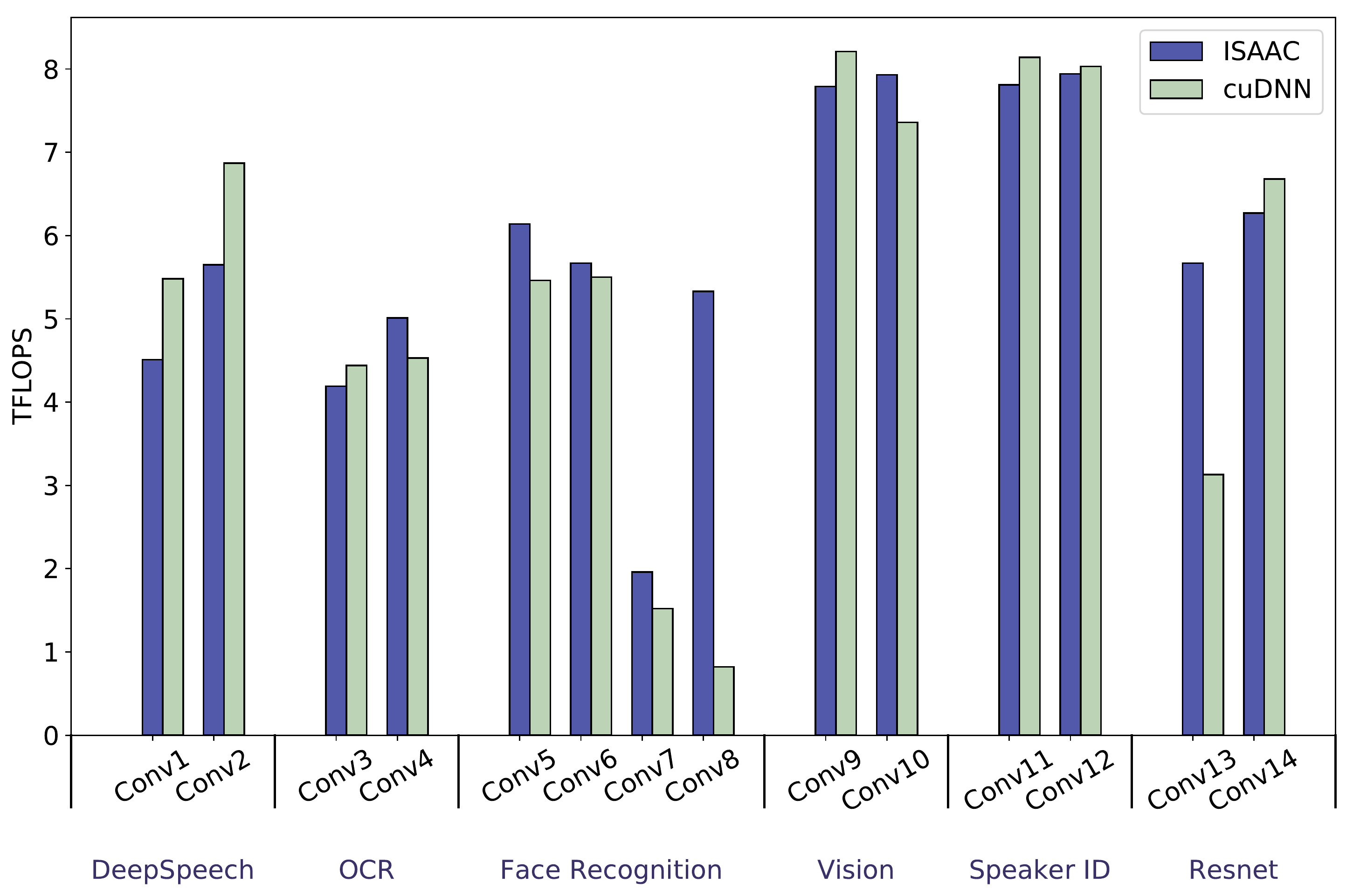}
	\caption{SCONV performance on the Tesla P100}
	\label{fig:bench:sconv:p100}
\end{figure}

\begin{figure}[h!]
	\includegraphics[width=.5\textwidth]{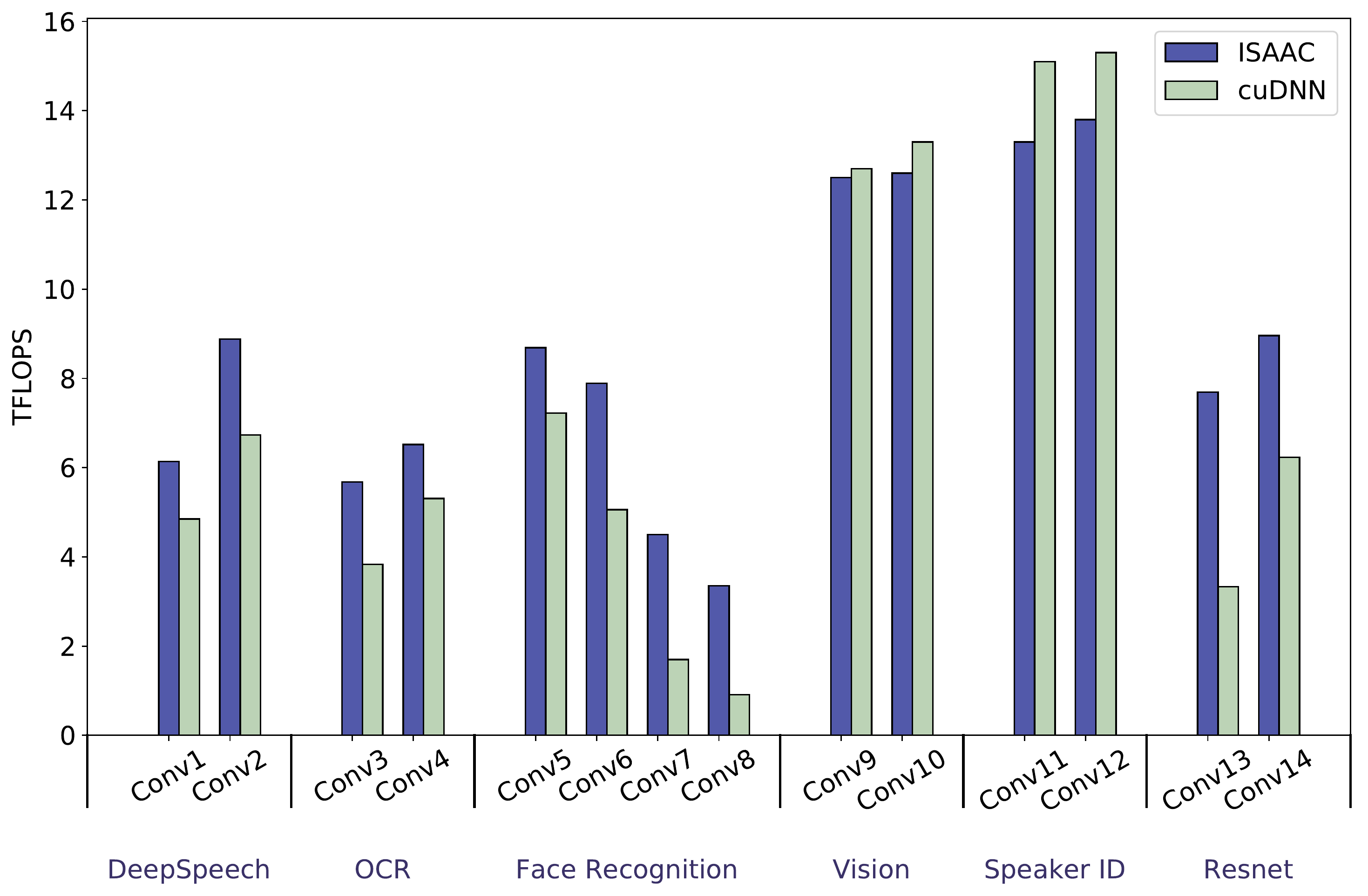}
	\caption{HCONV performance on the Tesla P100}
	\label{fig:bench:hconv:p100}
\end{figure}

Figure \ref{fig:bench:sconv:p100} and \ref{fig:bench:hconv:p100} show the performance of ISAAC for single- and half-precision convolutions, respectively. We observe large performance gains (more than $5\times$ for Conv8 and $70\%$ for Conv13) that we attribute to cuDNN's heuristics and kernels being tailored to Maxwell rather than Pascal.


As for HCONV, ISAAC's ability to easily support many tiling schemes result in almost consistently faster half-precision convolution routines than cuDNN.

\section{Analysis}
\label{sec:analysis}
The encouraging results shown in Section \ref{sec:numerical-benchmarks} beg for a thorough analysis of our system's performance: How exactly are such speed-ups achieved? What constitutes good parameter choices for our kernel generator? Why is PTX necessary to obtain good performance?

This section addresses these three questions, in order. We believe that the resulting insights could help library developers enhance existing software, like cuBLAS, which rely on a small set of statically generated kernels.

\subsection{DeepBench (Forward)}
Our previous benchmarks showed that, even in the presence of optimal kernel selection heuristics, cuBLAS could be up to 2x slower than ISAAC in single-precision, and 3x in half-precision. In order to explain why this is the case, this subsection provides a detailed comparison of ISAAC and cuBLAS's best kernel when (M, N, K) = (2560, 32, 2560), for the Tesla P100. 

The first thing to note is that this input configuration is only IO-bound under strong latency hiding assumptions: should the arithmetic operations not properly overlap with data-transfers, cycles will be lost and the achieved effective bandwidth reduced. Hence, it is unfortunate that cuBLAS only provides 64- and 128- way tiling along the N dimension, as it precludes the launch of enough warps to sustain high enough GPU occupancy (optimality is not a hard requirement since the problem is ideally still IO-bound).

\begin{table}[h!]
\centering
\begin{tabular}{|l|l|l|}
\hline
& \bf ISAAC & \bf cuBLAS\\
\hline
TFLOPS & 3.73 & 2.56\\
\hline
$M_L$ & 64 & 128\\
$N_L$ & 32 & 64\\
$K_L$ & 4 & 5\\
$P$ & 16 & 8\\
\hline
Shared Memory & 12.25kB & 12.25kB\\
Registers Count & 72 & 120\\
Occupancy & 17\% & 10\%\\
L2 hit rate & 32\% & 24\%\\
\hline
\end{tabular}
\end{table}

The main problem of cuBLAS's best kernel is that it assigns a large number of (de facto counter-productive) threads to an unexisting portion ($64 \leq N < 128$) of the result matrix. This has two adverse effects, which conjunctly explain the relatively bad performance we observed:
\begin{enumerate}
\item By using smaller tiling factors, ISAAC can decrease register/shared memory pressure, resulting in higher occupancy and therefore better latency hiding.
\item When higher occupancy no longer translates to improved performance, ISAAC (automatically) learns to use resources still available to instead pre-fetch more data into shared memory (i.e., larger $U$), resulting in better cache-hit rate (i.e., higher effective bandwidth).
\end{enumerate}
Reduction-spltting is, as mentioned previously, an alternative way to increase occupancy. Both ISAAC and cuBLAS use this method, although cuBLAS uses $K_L=1$.

\subsection{Kernel Selection}
The previous section may have given the reader a sense of what differentiates good from bad parameter values: (1) tile sizes should be small enough to guarantee high occupancy, but large enough to retain opportunities for ILP; (2) reduction splitting can be leveraged to further improve latency hiding, at the cost of diminished write bandwidth (via atomics) and/or additional shared memory usage, and (3) the pre-fetching factor $U$ can be increased to improve effective bandwidth when higher occupancy is no longer beneficial.

In order to better comprehend these trade-offs, Table \ref{tab:parameterization-choices} shows the parameterization choices made by ISAAC for the aforementioned problem sizes.

\begin{table}[h!]
\resizebox{\columnwidth}{!}{
\begin{tabular}{|l|l|l|l|l|l|l|l|l|}
\hline
\bf Problem & $\mathbf{M_s}$ & $\mathbf{N_s}$ & $\mathbf{M_L}$ & $\mathbf{N_L}$ & \bf U & $\mathbf{K_s}$ & $\mathbf{K_L}$ & $\mathbf{K_G}$\\
\hline
LINPACK (512) & 2 & 8 & 32 & 32 & 8 & 1 & 1 & 1\\
LINPACK (2048) & 8 & 8 & 64 & 64 & 8 & 1 & 1 & 1\\
\hline
DeepBench-F (16) & 2 & 4 & 64 & 16 & 16 & 1 & 1 & 4\\
DeepBench-F (128) & 4 & 4 & 64 & 32 & 8 & 1 & 1 & 2\\
\hline
DeepBench-B (16) & 4 & 2 & 16 & 16 & 16 & 1 & 8 & 1\\
DeepBench-B (128)& 4 & 4 & 64 & 64& 8 & 1& 1& 4\\
\hline
ICA (32) & 2 & 4 & 32 & 32 & 8 & 1 & 4 & 32\\
ICA (256) & 4 & 4 & 32 & 64 & 8 & 1 & 1 & 8\\
\hline
LAPACK (896) & 8 & 4 & 64 & 64 & 8 & 1 & 1 & 1\\
LAPACK (4096) & 8 & 16 & 64 & 128 & 4 & 1 & 1 & 1\\
\hline
\end{tabular}
}
\caption{Parameterization choices of ISAAC}
\label{tab:parameterization-choices}
\end{table}
ISAAC seems to properly learn to make sensible choices for all cases considered: (1) it chooses smaller tiles for smaller problems, (2) always split deep reductions problems (the proper trade-off is found between $K_L>1$ increases resources usage and $K_G>1$ which decreases write bandwidth) and (3) decreases $U$ appropriately to save hardware resources when good cache efficiency is not very important (see LAPACK).

\subsection{Advantages of PTX}
The first iteration of our software used CUDA-C and OpenCL for code-generation, but it was deprecated as adding bounds-checking resulted in a $15-20\%$ performance loss. Switching to PTX reduced this overhead to $2\%$. This is because modern NVIDIA hardware implement a mechanism called ``predication``: each instruction is complemented with a binary mask that specifies which thread should or should not be active. This mechanism, which does not require any program counter modification and has virtually no latency, is exposed in PTX but not in CUDA C.

 \section{Conclusions} \label{sec:conclusions}
 In this paper, we have presented ISAAC, an open-source\footnote{\url{https://github.com/ptillet/isaac}} framework for input-aware auto-tuning. Our tool relies on a versatile code generator able to adapt a wide range of problem sizes. We presented parameterization techniques for GEMM and CONV, used a multi-layer perceptron to model their behavior, and showed that features transformation was necessary to achieve proper convergence. We demonstrated how this model could be used to perform kernel selection at runtime, when input characteristics are fixed. Finally, we evaluated and analyzed the performance of our framework on a large variety of practical problems, and observed up to 3x performance gains over assembly-optimized vendor libraries.

Still, we see several possible directions of future work. While the good performance of our system on square matrices suggests that there is little room for improvement in our kernel generation mechanisms, our performance model relies on a series of rather basic techniques. Data-generation could be improved using better generative modeling techniques (e.g., Markov random field), and more efforts could be spent tuning our regression network. Another valuable addition to our framework would be a more flexible front-end (possibly a Domain Specific Language) to allow its use on problems beyond GEMM and CONV.

 \section{Acknowledgments} \label{sec:acknowledgments}
 This work was supported by the National Science Foundation (IIS 1409097) and by IARPA (contract D16PC00002).

\bibliographystyle{acm}
\bibliography{references}
\end{document}